\documentclass[prl,aps,twocolumn,groupedaddress,floats,showpacs,final]{revtex4-1}

\usepackage{graphicx}
\usepackage{dcolumn}
\usepackage{bm}
\usepackage{color}
\definecolor{blue}{rgb}{0.3,0.3,0.9}

\begin{document}

\title{Role of Bose Statistics in Crystallization and Quantum Jamming}

\author {M. Boninsegni$^{1,2}$, L. Pollet$^2$, N. Prokof'ev $^{2,3,4}$ and B. Svistunov$^{3,4}$}

\affiliation {$^1$Department of Physics, University of Alberta, Edmonton, Alberta, Canada T6G 2G7}
\affiliation {$^2$Department of Physics and Arnold Sommerfeld Center for Theoretical Physics, Ludwig-Maximilians-Universit\"at M\"unchen, 80333 M\"unchen, Germany
}
\affiliation {$^3$Department of Physics, University of Massachusetts, Amherst, Massachusetts 01003, USA}
\affiliation{$^4$Russian Research Center ``Kurchatov Institute", 123182 Moscow, Russia
}
\date{\today}
\begin{abstract}
Indistinguishability of particles is a crucial factor  destabilizing  crystalline order in Bose systems. 
We describe this  effect  in terms of damped quasi-particle modes and in the dual language of Feynman paths,
and illustrate it by first-principle  simulations of dipolar bosons and bulk condensed $^4$He. The first  major implication is that, contrary to conventional wisdom, zero-point motion alone  cannot prevent $^4$He crystallization at  near zero pressure. Secondly, Bose statistics leads to quantum jamming at finite temperature,  dramatically enhancing the  metastability of superfluid glasses. {Only} studies
of indistinguishable particles  can reliably address  these issues.
\end{abstract}


\maketitle
Among naturally occurring condensed matter systems, helium is the only known substance that escapes crystallization at low temperature ($T$), remaining a liquid all the way down to $T=0$  under the pressure of its own vapor. The standard argument  to explain the failure of liquid helium to form a crystal at low temperature  is based on its low atomic mass,  and consequently large  zero-point motion, and   the  weakness of the interatomic potential \cite{Lipson}.
\\ \indent 
Superficially, this contention appears plausible; indeed,  in the crystal phase of helium (stable at moderate pressure) the ratio between the zero-point atomic displacement  $u_0$ and the interatomic distance $a$,
an analog of Lindemann ratio ~\cite{lindemann} at zero temperature, is almost four times greater  than in all other known solids.
The standard argument makes no reference at all to Bose statistics, and in fact the assumption
is usually made that quantitatively accurate theoretical predictions of liquid-solid phase boundaries for Bose systems can be obtained by neglecting quantum statistics altogether, i.e., regarding particles as distinguishable. 
\\ \indent
Several reasons can be put forward to justify this approximation, which simplifies some   calculations, e.g., those based on Quantum Monte Carlo simulations
\cite{Coulomb}. 
For one thing, solid helium features many of the properties that are typical of other, less quantal solids, most notably an exponentially small energy scale (nearly five orders of magnitude smaller than vibration energies)
for tunneling exchange of two (and more) atoms \cite{Guyer,Mullin}. In other words, even in solid helium atoms are fairly  localized at their equilibrium (lattice) positions.
Quantitatively, the energy of a crystal  of $^4$He atoms is very nearly identical to  that of a crystal made of  distinguishable particles (henceforth referred to as {\it boltzmannons}) with 
the same mass and interatomic potential  of $^4$He atoms \cite{ceperley}.
\\ 
Thermal melting of  solids occurs when  fluctuations around equilibrium points, characterized by the r.m.s. excursion $u_T$, reach a certain fraction of the interatomic distance $a$,
that is when the Lindemann ratio $u_T/a$ is large enough (for most materials, this means above $\sim$ 0.1). This effect has little to do with quantum statistics, and occurs in the same way in a system of
boltzmannons.
\\ 
Finally, as was first noted by Feynman, there is absolutely no difference between the ground state wave function of an assembly of bosons, and one of correponding distinguishable particles.
For, the bosonic ground state wave function is nodeless, and boltzmannons are allowed to be in the same state by the symmetry of the Hamiltonian with respect to  particle label \cite{note}.
\\ \indent
In this Article, we argue that any attempt to determine the phase diagram of a Bose system  neglecting quantum statistics  is fundamentally flawed, prone to both large quantitative and qualitative errors.  Specifically, we show that, unlike their Bose counterpart,
distinguishable particle systems undergo thermo-crystallization above a liquid ground state, in a wide
region of the phase diagram, as a result of a mechanism akin to that of the Pomeranchuk effect in $^3$He.
We establish this conclusion through numerical simulations, of which we show results for two representative Bose systems, namely a two-dimensional assembly of dipoles interacting purely repulsively, and three-dimensional  condensed $^4$He. We furnish strong evidence that,  contrary to a widely held belief,  over a broad range of parameters  the destabilization of a crystalline phase is triggered by effects
related to Bose statistics, i.e., exchanges of indistinguishable particles, {\it not} by the conceptually distinct zero-point motion alone. In particular, we predict that $^4$He would {\it not} escape crystallization at low pressure and finite  temperature, were $^4$He atoms truly ``distinguishable''.
\\ \indent
We provide a simple theoretical framework for thermo-crystallization, making use of two alternative, complementary descriptions; 
the first is based on free energy considerations for a system of damped quasi-particles,
the second is formulated in the dual language of Feynman paths in imaginary time 
(i.e., world lines).
\\
A far-reaching implication of the importance of quantum exchanges, is the existence 
of a superfluid glass (superglass)
metastable phase, with long lifetime set by the entanglement of particle world lines involved
in macroscopic exchange cycles. Quantum jamming can take place in a Bose system, due to resilient  entanglement of indistinguishable particles, an effect that is entirely missed in studies of boltzmannons. In a Feynman path-integral picture, one can think of entangled world lines that  cannot be easily disentangled through a series of single-particle displacements.
\\ \indent
We begin by showing two examples of thermo-crystallization of boltzmannons, observed 
in computer simulations over a broad  range of thermodynamic parameters. Specifically, 
we describe simulation results for two
representative Bose systems with purely repulsive and Lennard-Jones type interaction potentials.
First, we consider a two-dimensional  assembly of bosons interacting via a  pairwise, purely repulsive dipolar potential $v(r)=D/r^3$, where $D$ is a constant. Such a many-body system can be realized in the laboratory by means of cold polar molecules \cite{pfau}, and its ground state phase diagram has been the subject of several  theoretical studies \cite{zoller,boronat}.
The Hamiltonian of the system in reduced units is
\begin{equation}\label{mod}
\hat{\mathcal{H}}=-\frac{1}{2}\sum_{i=1}^N \nabla_{i}^2+  \sum_{i>j} \frac{1}{r_{ij}^3},
\end{equation}
where $r_{ij}$ is the distance between particles $i$ and $j$.
The unit of length is $r_\circ\equiv mD$ (here and below $\hbar$=$k_B$=1), while $\epsilon_\circ \equiv D/r_\circ^3\equiv 1/mr_\circ^2$ is the energy unit. The relevant thermodynamic parameter, besides the temperature, is the  density $\rho =1/a^2$.
The system is enclosed in a cell with periodic boundary conditions and simulated
with the worm algorithm in the path-integral representation \cite{worm,worm2}.
\begin{figure}
\vspace*{-1.0cm}
\hspace*{-0.6cm} \includegraphics[scale=0.2,angle=0]{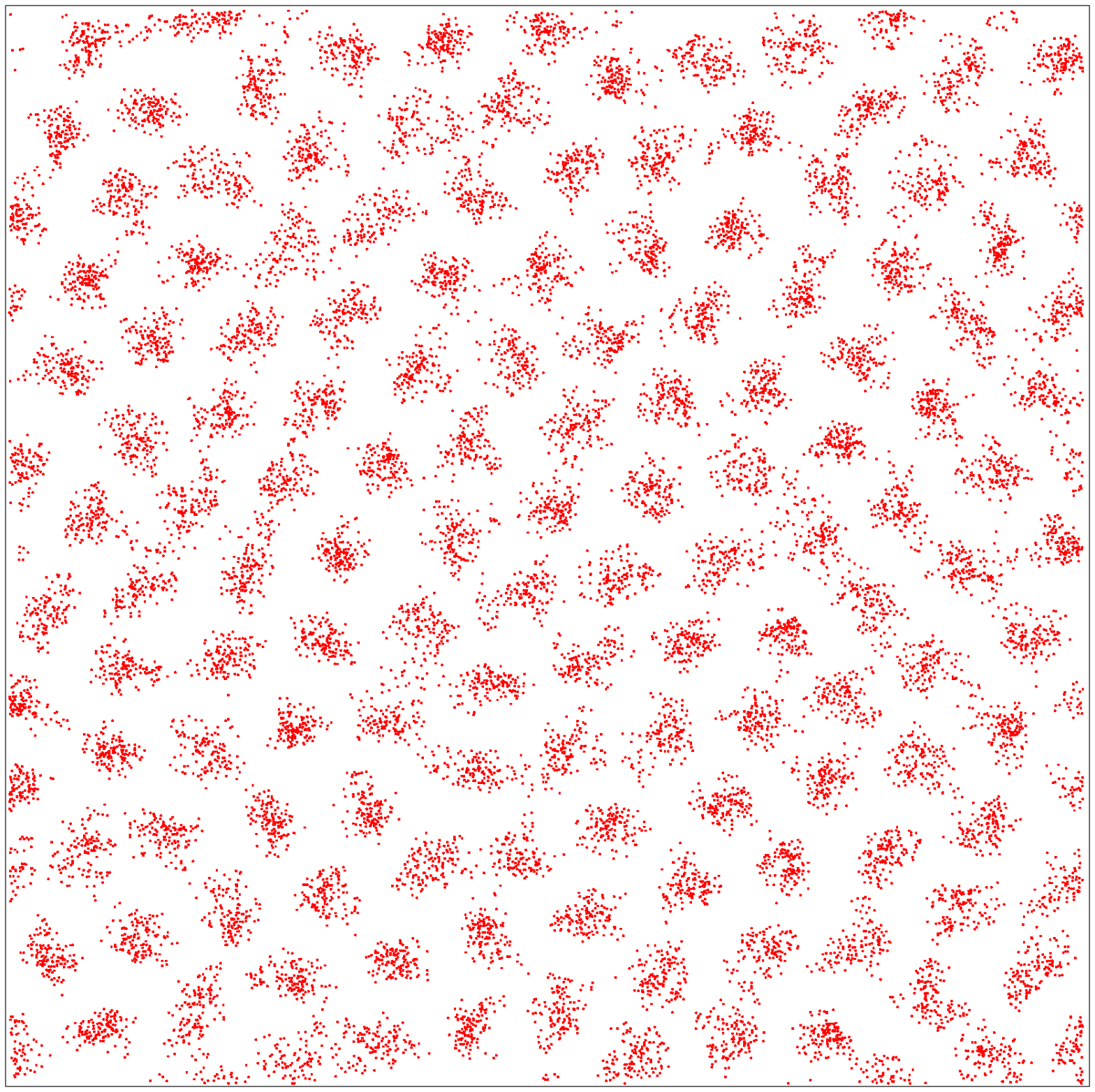}
\hspace*{-0.3cm} \includegraphics[scale=0.21,angle=0]{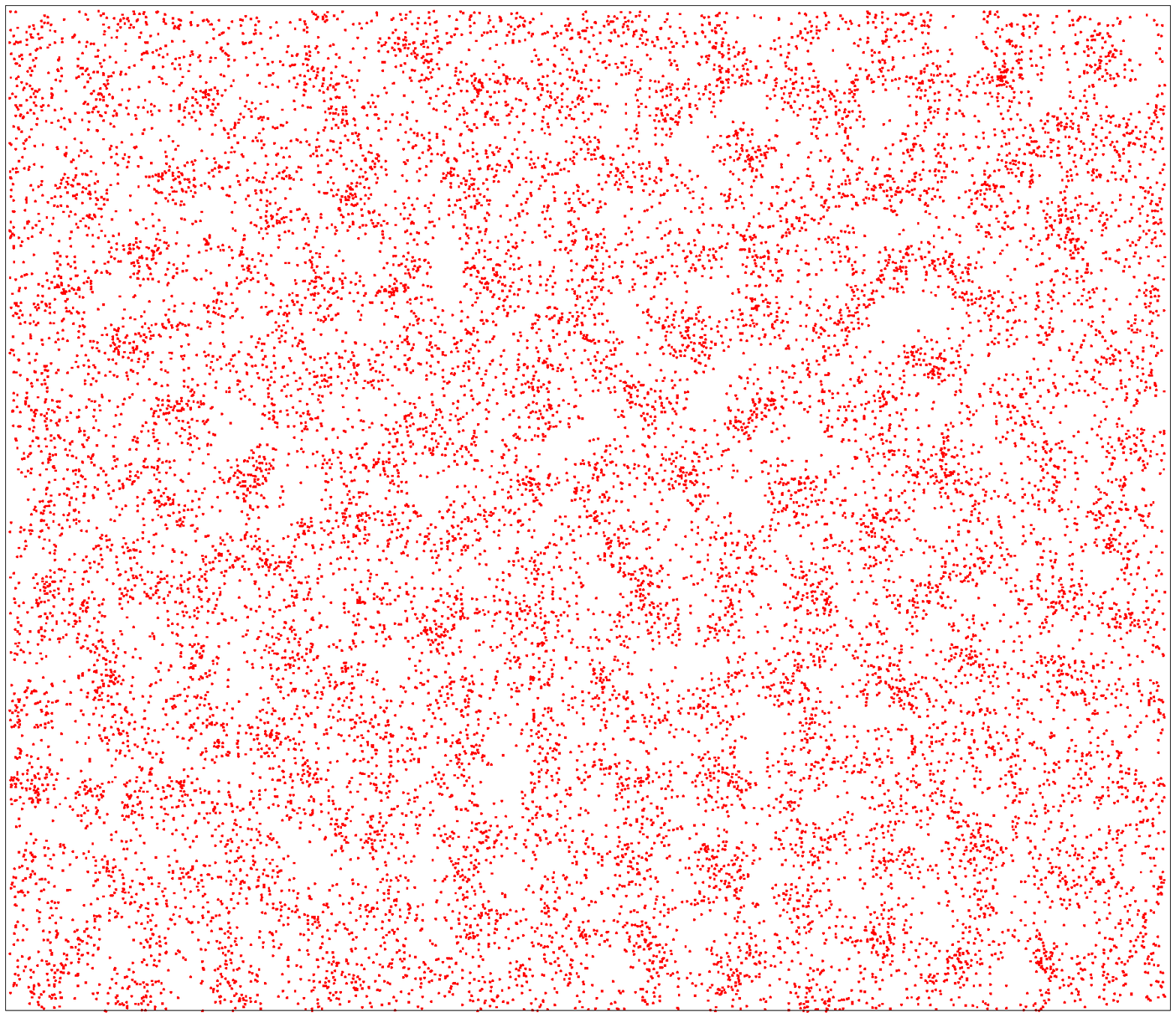}
\vspace*{-0.5cm}
\caption{\label{compare} Snapshots of  particle worldlines, projected onto the $xy$ plane, for a two-dimensional system of dipolar bosons, at temperature $T = 200\ \epsilon_\circ$ and interparticle distance $a=0.067$ (see text).
The size of each fuzzy `cloud' is a rough measure of quantum delocalization associated with zero-point motion.
Left panel shows an equilibrium configuration of boltzmannons,  right panel shows one 
of bosons, in which worldlines entangle. 
Simulation cell geometries and initial particle arrangements were purposefully chosen to favor the competing (i.e., non-equilibrium)  phase.}
\end{figure}

Figure \ref{compare} shows simulation snapshots of equilibrium particle worldlines, projected onto the $xy$ plane, for a system
of $N$=144 particles, at $T$=200 $\epsilon_\circ$ and $a=0.067$.
Although the simulation for boltzmannons is started with particles at random positions inside a square cell, the system {\it spontaneously} forms an ordered, triangular arrangement, clearly identifiable in the left panel of Fig. \ref{compare}. Fuzzy ``clouds" provide a rough measure of quantum delocalization arising from zero-point motion which is significant in this solid, as quantitatively expressed by the large value ($\sim$ 0.37) of the Lindemann ratio, far above the conventional threshold for (thermal) melting, and close to that of the bcc $^3$He \cite{glyde}.
We have observed crystallization of boltzmannons down to a temperature $T=50\ \epsilon_\circ$.
\\ \indent
The physical situation is qualitatively different if Bose statistics is taken fully into account, as shown in the right panel of Fig. \ref{compare}. In this case, although the system is initially prepared on a regular triangular lattice (hence the rectangular simulation cell), particle world lines entangle, and the crystalline order is destabilized in favor of a disordered, superfluid phase, with a value of the superfluid fraction close to unity \cite{noteadded}.
\\ \indent
We have observed the same behavior in a simulation of  three-dimensional condensed $^4$He, at a density
$\rho$ = 0.0248 \AA$^{-3}$ and temperature $T$=0.5 K (we made use of the accurate Aziz pair potential \cite{aziz}). The equilibrium thermodynamic phase in this case is a superfluid, and that is what we observe in a simulation of 108 {\it indistinguishable}  $^4$He atoms, i.e., the system quickly melts, if particles are initially arranged in a regular {\it hcp} crystal. However, in simulations in which exchanges cycles of particles are inhibited,  the system forms a {\it hcp} crystal, albeit with large atomic zero-point oscillations around lattice sites, even when atoms are initially placed at random positions. The difference in energy between the crystalline phase of distinguishable helium atoms, and the actual superfluid phase of $^4$He, is of the order of 0.5 K; exchanges contribute to lowering the atomic kinetic energy by approximately 1 K,  offsetting the potential energy increase due to the loss of crystalline order.
\\ \indent
In order to develop some appreciation for the importance of  thermo-crystallization, and how one cannot  ``sweep it under the rug", it is worth mentioning that $\rho$ = 0.0248 \AA$^{-3}$ in the {\it hcp} solid   is very close to zero pressure, as we verified by direct calculation (see also Ref. ~\cite{stress}). Thus, it is because of Bose statistics that the equilibrium thermodynamic phase of condensed helium at $T$=0.5 K is a superfluid, rather than a crystal.  By contrast, boltzmannons are capable of forming a solid phase at densities (and pressures) where vacancy and interstitial gaps are found to close, near zero pressure.
Our discussion is consistent with recent studies of boltzmannons with Coulomb interactions \cite{Coulomb}, which observe the effect of thermo-crystallization, with a slope linear in $T$ for the transition line between the liquid and the solid in the phase diagram.
\\ \indent
In order to arrive at a semi-quantitative theory of the thermo-crystallization phenomenon, we note that
dominant excitations in a system of Boltzmannons at low temperature are radically different
from sound waves in bosonic liquids and solids. If one considers a single particle as distinguishable from all others (e.g., as in Feynman's original study of a single $^3$He impurity in superfluid $^4$He \cite{feynman}),  one may assume an impuriton excitation characterized by a
parabolic dispersion relation $\epsilon_k=k^2/2m^*$, with effective mass $m^*=m^*_{L,S}$ for liquid and solid
phases, respectively.  Impuriton type excitations with the same $\epsilon_k$ remain well-defined
in a system of boltzmannons, because their ground state is homogeneous due to particle
delocalization over the entire system volume. At finite temperature, delocalization of
particle labels over the de Broglie wavelength $\lambda_T=\sqrt {2\pi /m^*T}$
guarantees that unmodified impuriton excitations exist
at energies $\epsilon \gg T$; their inverse life-time $\tau^{-1} \sim \sqrt{T \epsilon}$
can be estimated from the condition of traveling a distance $\lambda_T$. For thermodynamic applications,
however, damping rates comparable to the typical quasi-particle energy do not change scaling of energy
and free energy with temperature, and thus semi-quantitative results can be obtained by considering
a gas of $N$ distinguishable quasi-particles with dispersion relation $\epsilon_k$.
\\ \indent
This picture applies to both  liquid and crystal, but for the latter one  should take into account that
$\epsilon_k$ is  a tight-binding dispersion relation, with an exponentially small bandwidth
$W$, set by tunneling exchange of particles. Thus, at $T \ll W$ we are dealing
with an effective mass $m^*_S \sim 2d/Wa^2$, orders of magnitude larger than $m^*_L$ (in helium,
$m^*_S/m^*_L \sim 10^4$).
At $T > W$, the thermal part of free energy of boltzmannons (and any other multi-component system)
is in leading order purely entropic: $\Delta F_S = F(T)-F(0)\approx -TS$, where $S=\ln (N!) $ (in general,
$e^S$ is the number of distinguishable permutations in the $N$-particle system).
When bosons are replaced with boltzmannons, both the crystal and the liquid acquire negative contributions to
their free energies, but $\Delta F_S$ is larger in magnitude than $\Delta F_L$.
This enables a transition to a solid phase at finite temperature, even if the ground state is liquid.
The effect  of thermo-crystallization discussed here is similar to the
liquid-solid transition in $^3$He, which is driven by fast
spin entropy increase (up to  $\ln 2$ per particle) in the crystalline phase.
\begin{figure}
\vspace*{-0.7cm}
\includegraphics[scale=0.3,angle=0]{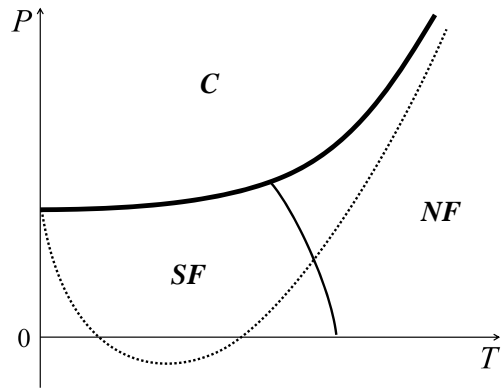}
\vspace*{-0.5cm}
\caption{\label{sketch} Schematic liquid-solid phase diagram for Lennard-Jones type Bose systems.
Solid lines separate the normal fluid (NF), superfluid (SF) and crystalline (C) phases.
The dashed line shows the position of the crystal-liquid boundary
if quantum exchanges are neglected. In case of $^4$He mass and potential, distinguishable particles predict
a solid phase around $T=0.5$~K. Phases at $P<0$ are metastable as vapor (excluded here for clarity)
is the equilibrium state at finite temperature and zero pressure.
Note that in the $T\to 0$ limit the descriptions in terms of distinguishable and Bose particles become equivalent~\cite{note}. }
\end{figure}

Quantitatively, using standard expressions for the ideal gas,
\begin{equation}\label{mod}
F_{L}- F_{S}\, \sim \, \Delta E(0) + \frac{d}{2} NT\,  \ln {m^*_S\over m^*_L } \qquad (T \ll W) \, ,
\label{f1}
\end{equation}
\begin{equation}\label{mod}
F_{L}- F_{S} \sim \Delta E(0) + \frac{d}{2} NT\,  \ln {T_{\rm deg}\over T } \quad (W \ll T \ll T_{\rm deg} ) ,
\label{f2}
\end{equation}
where $d$ is the dimensionality,  
$\Delta E(0) = E_L(0)-E_S(0)$ is the energy difference in the ground state and
$T_{\rm deg} \sim  2 \pi n^{2/d}/m^*_L $ is an estimate of the liquid degeneracy temperature.
Note the finite slope of the liquid-solid interface at low temperature,
$T_c \propto (-\Delta E(0)/N)/ \ln (m^*_S/ m^*_L )$, and the fact that this slope can be quite small
due to exponentially large effective mass ratio. At $T>T_{\rm deg}$ both phases fully realize their
$\ln (N!)$ entropies and effects of quantum statistics become unimportant.
The resulting phase diagram is illustrated in Fig.~\ref{sketch}. The reentrant behavior 
will take place in {\it any} system of distinguishable particles close to the liquid-solid 
quantum critical point. The dashed curve may or not extend to regions of negative pressure.
In order not to overload the discussion we focus on the liquid-solid transition, 
i.e., we exclude vapor, and gloss over possible hexatic (in 2D) and supersolid phases which
may exist for certain systems.
\\ \indent
In the path-integral language, the difference between bosons and boltzmannons is exclusively
in the nature of periodic boundary conditions in imaginary time $\tau \in (0, 1/T)$, in that
world lines for distinguishable particles are not allowed to form exchange cycles.
Feynman's theorem for the ground state is understood as strongly fluctuating paths, such that looking at any
imaginary time scale one is unable to determine whether the worldlines would eventually form exchange cycles or not: at $\tau \ll 1/T$ the system can well be mapped to a system with open boundary conditions in imaginary time. From this picture,
we deduce that propagation of the impuriton worldline in imaginary time $\tau \ll 1/T$,
described by the Matsubara Green's function, is the same in bosonic and boltzmannon systems,
in agreement with the previous conclusion. Interestingly, thermo-crystallization is now linked to the
increased ``bending" energy (action, to be more precise) for highly entangled worldlines to
satisfy boundary conditions. Whereas bosonic worldlines reconnect on each other and
form large (macroscopic) exchange cycles,  the distinguishable lines have to return to their
original positions. This puts liquid worldline configurations of boltzmannons at a disadvantage
with respect to their bosonic counterpart.
In the vicinity of the superfluid-solid transition, {\it long} exchanges of indistinguishable particles
(comprising a finite fraction of all particles in the system) play a crucial role in stabilizing the equilibrium superfluid phase. Hence, neglecting quantum exchanges results in an incorrect characterization
of the physics of the system.
\begin{figure}
\vspace*{-0.5cm}
\hspace*{-0.6cm} \includegraphics[scale=0.3, angle=0]{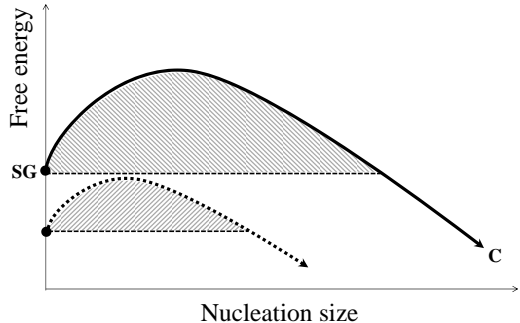}
\vspace*{-0.9cm}
\caption{\label{thermonuc} An illustration of the nucleation process leading to large crystalline seeds.
Compared to their bosonic counterpart (upper part), the distinguishable particles (lower part) have a lower free-energy nucleation barrier due to the large negative free energy shift of the crystalline phase.}
\end{figure}

Long exchange cycles are also crucial for determining the lifetime and metastability of superglasses.  A plausible order-of-magnitude estimate of the
lifetime of a superglass phase of $^4$He, places it at least in the millisecond range \cite{superglass},
which should allow for its experimental observation.
Classical jamming is understood as a physical process wherein the system enters a structurally disordered state, with severely restricted motion of individual particles, which cannot
easily rearrange into a more energetically favorable configuration. In its extreme form,
it leads to the formation of classical glasses. An intuitive characterization of the superglass phase consists of   quantum jamming
of highly entangled particle world lines.
On the one hand, this state has frozen structural
disorder on a microscopic scale, on the other hand it can support dissipation-less flow of its own particles, if macroscopic exchanges extend throughout the whole system.
In order for the system to find its equilibrium lower-energy ordered (and insulating) configurations, world lines have to disentangle from macroscopic exchange cycles, and this can be expected to involve rare multi-particle exchanges, i.e. the nucleation of the crystal phase proceeds through the multi-particle seed.
At finite temperature, it is reasonable to expect that the leading channel
should be thermal (rather than quantum) nucleation.  In this case, illustrated in Fig.~\ref{thermonuc}, the free energy barrier is higher for indistinguishable particles, rendering the probability of nucleation dramatically lower
compared to that of boltzmannons.
Thus, effects of Bose statistics are instrumental in conferring enhanced stability to both liquids and superglasses.
Studies of boltzmannons at finite temperature miss the quantum jamming effect of long exchange cycles, and
can not therefore be used to address existence and stability of superglasses~\cite{biroli}.
\\ \indent
In conclusion, we showed that indistinguishability of particles has profound effects on the phase diagram of Bose systems, $^4$He being the most obvious case. The inclusion of exchanges of indistinguishable particles in the formalism is crucial, in order to obtain correct phase transition lines, zero-point motion alone being insufficient at finite temperature, contrary to a claim routinely made even in textbooks \cite{Lipson}. For superglass phases, particle entanglement
caused by long exchange cycles greatly enhances its lifetime against crystallization, an effect
that cannot be captured by studies in which Bose statistics is neglected.
\\ \indent
This work was supported by the Natural Science and Engineering Research Council of Canada under research grant G121210893 and by the National Science Foundation under grant PHY-1005543.


\begin{thebibliography}{99}
\bibitem{note} The $T\to 0$ limit  for distinguishable particles
is approached for a system with finite volume $V \sim L^d$, for temperatures lower than  the corresponding discretization energy gap
$\Delta E \sim \hbar^2 /mL^2$.
\bibitem{noteadded}
Obviously, although we only show here a qualitative, visually suggestive image such as Fig. \ref{compare} for illustrative purposes,  liquid and solid phases have been quantitatively identified in this work through the computation of physical observables such as the pair correlation function.
\end{thebibliography}
\end{document}